\documentstyle[12pt,aasms4]{article}

\lefthead{Raga \& Noriega-Crespo}
\righthead{}

\def \yskip{\penalty-50\vskip3pt plus 3pt minus 2pt}
\def \pp{\par \yskip \noindent \hangindent .4in \hangafter 1}
\def \abc#1#2#3#4 {\pp#1, {\sl#2}, {\bf#3}, #4}
\def \blank {\lower 5pt\hbox to 0.75in{\hrulefill}}

\newfont{\rten}{cmr10}

\begin{document}

\title{A 3-mode, variable velocity jet model for HH 34}

\author{A. Raga\altaffilmark{1} and A. Noriega-Crespo\altaffilmark{2}
\altaffiltext{1}{Instituto de Astronom\'\i a,
UNAM, Ap. 70-264, 04510 M\'exico, D. F., M\'exico}
\altaffiltext{2}{Infrared Processing and Analysis Center,
CalTech-JPL, Pasadena, CA 91125, USA}}

\begin{abstract}
Variable ejection velocity jet models can qualitatively explain the
appearance of successive working surfaces in Herbig-Haro (HH) jets.
This paper presents an attempt to explore which features of the
HH~34 jet can indeed be reproduced by such a model. From previously
published data on this object, we find evidence for the existence
of a 3-mode ejection velocity variability, and then explore the
implications of such a variability. From simple, analytic
considerations it is possible to
show that the longer period modes produce a modulation on the shorter
period modes, resulting in the formation of ``trains'' of multiple
knots. The knots observed close to the source of HH~34 could correspond to
such a structure. Finally, a numerical simulation with the ejection velocity
variability deduced from the HH~34 data is computed. This numerical
simulation shows a quite remarkable resemblance with the observed
properties of the HH~34 jet.
\end{abstract}


\keywords{Stars: Mass-loss --- Hydrodynamics --- Shock Waves --- Star
Formation --- HH~34}

\section{Introduction}

The Herbig-Haro (HH) object HH~34 presents a remarkable, jet-like
structure that has been studied extensively over the years.
The first extensive studies were carried out by Reipurth et al. (1986)
and B\"uhrke et al. (1988), presenting radial velocities, line ratios
and images in different emission lines. Later work on this object
includes high (Heathcote \& Reipurth 1992) and low (Morse et al.
1993) resolution spectroscopy, Fabry-Perot imaging interferometry
(Morse et al. 1992) and proper motion measurements (Eisl\"offel \&
Mundt 1992; Heathcote \& Reipurth 1992).

There has also been a substantial amount of work in using deconvolutions
to try to resolve the knots along the HH 34 jet (Raga \& Mateo 1988;
B\"uhrke et al. 1988; Raga et al. 1991). These results have been
superseded by the high resolution images of Reipurth \& Heathcote
(1992, who present images obtained with the ESO New Technology Telescope)
and by the HST images (still not published in detail, but discussed
in a partial way by Ray et al. 1996 and Reipurth et al. 1997).
Unfortunately, these HST images have a rather low signal-to-noise
ratio, complicating possible comparisons with theoretical models.

Recently, it has been shown that the HH~34 flow is the central
part of a ``superjet'' (Bally \& Devine 1994; Devine 1997; Eisl\"offel
\& Mundt 1997; Devine et al. 1997) extending over a total distance of
$\sim 3$~pc. This result puts the previous known facts of the HH~34
system in a totally new perspective. HH~34S (i.~e., HH~34 itself,
as listed in the catalog of Herbig 1974) has now been demoted from
being the head of a jet, to being one of several ``internal working
surfaces'' inside the blueshifted outflow lobe.

Finally, we should note that the source of HH~34 is visible optically.
Rodr\'\i guez \& Reipurth (1996) have detected a continuum VLA source
that coincides (to within $\sim 1''$) with the optically detected star.
This is an important difference between HH~34 and the other well studied
objects HH~46/47 and HH~111, which are seen to emerge from dense
cores which obscure the central source.

From the theoretical point of view, efforts have been made to compute
3/2-D bow shock models for reproducing images (Raga 1986) and line
profiles (Morse et al. 1992) of HH~34S. Raga \& Noriega-Crespo (1992)
tried to apply such models for explaining the emission from the
successive knots in the region between the source and 34S. This
application of bow shock models to the aligned knots is based on
the analytic work of Raga et al. (1990, see also Raga \& Kofman 1992;
Kofman \& Raga 1992), who suggested that these knots could correspond
to internal working surfaces resulting from an ejection velocity
variability. This idea was pursued with 1D numerical simulations
by Hartigan \& Raymond (1993) and with cylindrically symmetric and 3D
simulations by a number of authors (see, e.~g., Stone \& Norman 1993;
Gouveia dal Pino \& Benz 1994; Biro \& Raga 1994; Smith et al. 1997b),
and with high resolution simulations of single internal working
surfaces (Falle \& Raga 1993, 1995; Biro 1996).
The interaction between successive working surfaces has also been studied.
The statistical properties of the flow pattern resulting from many
collisions between thin working surfaces were described by Raga (1992),
and the shock reflection processes resulting from widening working
surfaces (which get to touch each other) were described by Smith et
al. (1997a).

The present paper discusses an application of models of jets from
variable velocity sources to the case of the HH~34 jet. In particular,
the region between the source and HH~34S itself is modeled, and no
attempt is made to model the ``superjet'' beyond HH~34S (though
the observations of this region are discussed in a partial way).
The paper is organized as follows. Section 2 summarizes the results from
analytic working surface models, and their implications for HH~34.
Section 3 discusses previously obtained
observational results for the knots along
the HH~34 jet, and the derivation of the parameters of the source
variability that can be carried out using the observations and the
results from the analytic models. Section 4 describes the
results of a radiative, axisymmetric numerical simulation with
the deduced parameters for the source variability. Finally, the results
from this simulation are compared with observations of HH~34 in section 5.

\section{Analytic working surface models}

Let us now discuss some simple, analytic considerations that
can be made about the formation and propagation of internal working
surfaces. First, we derive simple expressions for the distances
from the source at which working surfaces form, and at which they
achieve maximum shock strengths (section 2.1). We then summarize
some of the results of the asymptotic regime of large distances
from the source (section 2.2). Finally, we discuss the qualitative
features of jets from sources with multi-mode velocity variabilities
(section 2.3). The qualitative implications of these results for
the case of HH~34 are discussed.

\subsection{The formation of internal working surfaces}

Let us consider a jet from a source with a sinusoidal ejection velocity
variability~:

$$u_0(t)=v_0+v_a\,\sin \omega_at\,,\eqno(1)$$

\noindent where $v_0$ and $v_a$ are constants, and $\omega_a=2\pi/
\tau_a$ is the frequency of the variability (with $\tau_a$ being the
period). For the $v_a\ll v_0$ case, from the results of Raga et al.
(1990) it is straightforward to show that the working surfaces resulting
from this ejection velocity variability are formed at a distance
from the source~:

$$x_c={v_0\over v_a}{\Delta x\over 2\pi}\,,\eqno(2)$$

\noindent where

$$\Delta x=v_0\tau_a\,,\eqno(3)$$

\noindent is the separation between successive working surfaces.
The velocity jump $\Delta v$ across the working surface has a low (basically
sonic) value close to the point in which the working surface is formed
(i.~e., at $x=x_c$), and rapidly grows until it reaches a maximum
$\Delta v\approx 2v_a$ at a distance~:

$$x_m\approx {\pi\over 2} x_c\,,\eqno(4)$$

\noindent from the source.

These equations have direct implications for the formation of knots
along HH jets. For the aligned knots radiating away from the sources of
HH jets to be observable in optical atomic or ionic lines, they need to
have shocks with velocities at least of order $\sim 10$~km~s$^{-1}$.
From this, we conclude that we need to have $v_a\approx 10$~km~s$^{-1}$
or larger. The velocities of HH jets indicate that $v_0\sim 100$~km~s$^{-1}$.
Inserting these values into equation (2), we conclude that the working
surfaces have to form at a distance $x_c\sim \Delta x$ from the source.
Also, from equation (4) we see that the maximum shock strengths
are also attained at a distance from the source which is comparable
to the knot spacing. As we discuss in section 3, this general result
does not appear to agree with the observed properties of the HH~34 jet.

\subsection{The asymptotic regime of large distances from the source}

In this section, we summarize some of the results of the asymptotic
regime (of large distances from the source) studied by Kofman \& Raga
(1992) and Raga \& Kofman (1992). At large distances from the source
(i.~e., at distances several times larger than the knot separation
$\Delta x$), the velocity jump across the working surface is given
by~:

$$\Delta v={{v_0}^2\tau_a\over x}\,,\eqno(5)$$

\noindent where $x$ is the distance from the source to the working
surface, and $v_0$ and $\tau_a$ are defined in equation (1). Using
a power law fit to the plane-parallel shock models of Hartigan et al.
(1997), it is also possible to show that the H$\alpha$ luminosity
of the knots scales with distance as~:

$$L_{H\alpha}\propto x^{-4.8}\,.\eqno(6)$$

\noindent These results are used to interpret the observations of
HH~34 in section 3.

\subsection{Two-mode interactions}

Finally, let us consider the more complex case of a two-mode
ejection velocity variability~:

$$u_0(t)=v_0+v_s\,\sin \omega_st+v_f\,\sin \omega_ft\,,\eqno(7)$$

\noindent with a ``slow'' variability of amplitude $v_s$ and
period $\tau_s=2\pi/\omega_s$, and a ``fast'' variability of
amplitude $v_f\ll v_s$ and period $\tau_f=2\pi/\omega_f\ll \tau_s$.

The ejection velocity variability described by equation (7) is shown in 
Figure~1 for two sets of parameters. From this figure, it is clear that
the values of the difference $\Delta u_f$ between the minima and
following maxima of the ``fast'' variability (see figure 1)
are affected by the presence
of the ``slow'' variability. From an analysis of equation (7),
it is straightforward to show that~:

$$\Delta u_f(t)=2v_f\left\{\sqrt{1-\eta^2}+\eta\,\arcsin
\left(\sqrt{1-\eta^2}\right)\right\}\,;\,\,\,\eta\le 0$$

$$\Delta u_f(t)=2v_f\left\{\sqrt{1-\eta^2}+\eta\,\left[\pi-\arcsin
\left(\sqrt{1-\eta^2}\right)\right]\right\}\,;\,\,\,\eta>0\,,\eqno(8)$$

\noindent where the first branch of the arc sine is taken, and

$$\eta(t)=\eta_0\,\cos \omega_s t\,,\,\,\,{\rm with}\,\,\, \eta_0=
{v_s\tau_f\over v_f\tau_s}\,.\eqno(9)$$

\noindent For $|\eta|\ll 1$, equation (8) takes the simple form~:

$$\Delta u_f\approx 2\,v_f\,\left(1+{\pi\over 2}\eta\right)\,.\eqno(10)$$

From equation (8) or (10) it is possible to see that for $\eta=0$
(i. e., for a zero amplitude slow variability), we have $\Delta v_f=2v_f$.
During the time intervals in which the slow velocity variation is
rising (i. e., when $\eta>0$), the velocity jumps
associated with the fast variability are boosted upwards to
values $2v_f<\Delta u_f\leq 2\pi v_f$. The maximum value of $\Delta
u_f=2\pi v_f$ is attained for $\eta=1$ (see equation 8), and for $\eta>1$ the
minima and maxima associated with the fast variability actually disappear
(as illustrated in the second panel of figure 1).
Conversely, for the time intervals with $\eta<0$ (see equation 9),
we have values $0\leq \Delta u_f < 2v_f$, with the lower value
attained for $\eta\leq -1$ (see equation 8).

From this discussion, we see that a long period variability of the
source has the effect of ``modulating'' the shorter period variability.
The short period, ``fast mode'' will produce working surfaces of
increased strength for the time intervarls in which the ``slow mode''
variability gives a rising ejection velocity vs. time. Conversely, the
``fast mode'' produces working surfaces of decreased strength for the
time intervals in which the ``slow mode'' gives a negative ramp
of decreasing ejection velocity vs. time. This discussion strictly
applies for two modes with parameters such that the condition $\eta_0<1$
(see equation 9) is met.

In this way, the long period variability has the effect of modulating
the knots along the jet (i. e., the working surfaces produced by the
short period variability) into ``trains'' of knots of enhanced emission
travelling down the jet beam. These trains of knots have a length
of $\sim v_s \tau_s/2$. The knots along the HH~34 jet could in principle
correspond to such a train of knots resulting from the interaction
of two modes of the ejection velocity variability.

\section{Observations of HH~34}

\subsection{HST images}

We have reduced archival HST images of the HH~34 flow with the
standard, pipeline reduction procedure. These images have been
previously studied by Ray et al. (1996) and by Reipurth et al. (1997),
but have not yet been analyzed completely.

The region from the source out to knot L is shown in Figure 2. From this
figure we see that even though the region between the source and knot
E (located at an angular distance of $\approx 11''$ from the source)
is very faint, the jet is still detected. The emission of this region
appears to be somewhat narrower than the brighter region at distances
larger than $10''$ from the source (as pointed out by Ray et al. 1996).

An important question is why the region at distances $<10''$ from
the source is so faint. A possible explanation of course is that the
jet could be emerging from a high extinction region, which would extend
out to a distance of $\sim 10''$ from the source. The existence of
such a high extinciton region would also be a possible explanation for
the lack of an observable counterjet in the red-shifted lobe (out
to the red-shifted working surface HH~34N, see, e.~g., B\"uhrke et al.
1988).

However, if the region close to the HH~34 source were subjected to
high extinction, we would expect to see an apparent broadening of the
jet beam towards the source, as a result of the existence of an increasingly
strong scattered component. This broadening has been modeled (for the
case of a jet emerging from a stratified clump) by Feldman \& Raga
(1991), who predict a quite dramatic spatial broadening of the emission
as the observed intensity drops (as a result of increasing extinction and
dispersion). The effect predicted by Feldman \& Raga (1991) would be less
dramatic for the case of a jet travelling within a cavity which has
been opened up in the surrounding, stratified core.

The fact that in the HST images the jet emission becomes narrower
at distances $<10''$ from the source (see above and Ray et al. 1996)
therefore appears to rule out extinction as a possible explanation for
the low intensity of this region. From this argument we then conclude
that a model for the formation of knots in the HH~34 jet has to
explain this sudden brightening of the jet at $\approx 10''$ from the
source. As is discussed below, this point is of particular interest.

For the HH 34 knots we have a spatial velocity
$v_0\approx 200$~km~s$^{-1}$ (see section 3.2), an amplitude for
the velocity variability of $v_a\approx 20$~km~s$^{-1}$
(necessary to produce shocks which lead to observable emission),
and a separation $\Delta x\approx 2''$ between successive knots.
Then, from equation (2) we obtain $x_c\approx 3''$ and from equation
(3) $x_m\approx 5''$. So, we would expect the knots along the jet
to start brightening at $\sim 3''$ away from the source, peak in intensity
at $\sim 5''$, and decay for larger distances from the source.
Therefore, we cannot explain the $\sim 10''$ gap between the source and
the bright knots along the HH 34 jet with a single-mode, sinusoidal
source variability.

As we have described in section 2.3, two-mode interactions
produce trains of knots travelling down the jet beam. We tentatively
identify the HH 34 knots with such a train. In the following, we explore
in detail whether this interpretation is indeed appropriate for the knots
along the HH 34 jet.

\subsection{Proper motions, radial velocities and H$\alpha$ intensities}

In order to evaluate the properties of the HH 34 jet in a more quantitative
way, we now summarize the measurements that have been carried out of
proper motions, radial velocities and H$\alpha$ fluxes for the knots
along the jet. Figure 3 shows the spatial velocity and the luminosity
of the knots as a function of distance from the source. The values
shown in this figure are also tabulated in Table 1.

The distance from the source has been calculated assuming a distance
to HH 34 of 450 pc, and an orientation angle $\phi=30^\circ$ between
the axis of the outflow and the plane of the sky (there now seems
to be a partial consensus on the value of this angle, see, e.~g., Heathcote
\& Reipurth 1992; Morse et al. 1992; Raga et al. 1997a). The spatial
velocity of the knots has been calculated by deprojecting the
proper motion velocities measured by Heathcote \& Reipurth (1992)
(similar values are obtained by deprojecting the radial velocities),
except for the knots HH 87, 88 (belonging to the S lobe of the HH
34 superjet), for which we have deprojected the radial velocities
measured by Devine et al. (1997), as the proper motions measured
for these knots appear to have larger errors.

We have computed the H$\alpha$
luminosities of the knots by integrating inside circular apertures with
diameters chosen to include the emission of the successive knots,
and then carrying out a background subtraction by using fields to the two
sides of the outflow axis. For this calculation, we have used the
HST archival H$\alpha$ image (for knots E through HH 34S), and the
images of the superjet of Devine et al. (1997). A relative calibration
between the different images was obtained using the knots that
appear in several images. Finally, the measured relative intensities
were calibrated using the H$\alpha$ flux measured for HH~34S
by Morse et al. (1992).

From Figure 3 (also see table 1), we see that the spatial velocity
of the knots initially rises (from knot E to knot G), then decays
(from knot G to K) and finally rises monotonically (from knot O to
HH 34S). The HH 87, 87 objects (belonging to the HH 34 superjet)
show a substantially lower spatial velocity.

From this figure we also see that knots E-K and O have similar luminosities,
and that HH 34S is by far the brighter condensation. On the H$\alpha$
luminosity vs. position graph we have also plotted (with dashed lines,
see figure 3) the $L_{H\alpha}\propto x^{-4.8}$ law predicted from the
asymptotic solution for large distances from the source (see equation 6).
As pointed out by Raga \& Kofman (1992), we see that knots I-L approximately
fall on a single dashed line. However, all of the other knots fall
on different lines.

This result implies the following. The observed knots are formed at
a finite distance from the source (given by equation 2), rapidly
reach their maximum intensity (at the distance given by equation 4),
and then decay in intensity as they continue travelling away from
the source with the predicted $x^{-4.8}$ dependence. Looking at
Figure 4, we then see that by the time that the knots E-L get to
the present position of knot O, they will be fainter than the
present knot O by more than 2 orders of magnitude. We therefore
conclude that the variability giving rise to knots E-L is not
the same variability as the one that gave rise to knot O. Analogously,
we can argue that variabilities with different periods and amplitudes
are also necessary for producing HH 34S and HH 87, 88.

In this way, we come up with a picture in which different modes of
oscillation of the ejection velocity give rise to knots formed at
different distances from the source. As these knots travel away from
the source, they rapidly decay approximately following a $x^{-4.8}$
law. Given this rapid decay (and the existence of a detection limit
in the observations), the positions of
the observed knots have to be close (at least to an order of magnitude)
to the position at which they were formed.

The observed distribution
of knots therefore can be interpreted as evidence for the existence
of different modes of oscillation of the ejection velocity.
It is, however, not clear how to deduce the properties (period
and amplitude) of these different modes. In the following two subsections
we describe possible ways of deducing these properties for the knots
of the outflow from the source out to HH~34S.

\subsection{The HH~34S working surface}

Let us now consider what ejection velocity variability is needed for
producing the HH~34S working surface. Figure 4 is a schematic representation
of the [S II] 6717+31 position-velocity diagram of HH~34S presented
by Reipurth (1989a) and Heathcote \& Reipurth (1992). From this figure,
we see that the material feeding into the working surface from the
upstream direction has a (de-projected) velocity $v_1=344$~km~s$^{-1}$,
and that the working surface is overruning downstream material which
is propagating away from the source at a velocity $v_2=214$~km~s$^{-1}$.
These two velocities straddle the value of the spatial velocity of
HH~34S deduced from its proper motion (see table 1).

Using these velocities and the distance $x_{34}={8.0\times 10^{17}}$~cm
from the source to HH 34S, it is possible to compute the dynamical
timescales

$$t_1={x_{34}\over v_1}=738\,{\rm yr}\,,\eqno(11)$$

$$t_2={x_{34}\over v_2}=1187\,{\rm yr}\,,\eqno(12)$$

\noindent of the material upstream and downstream of 34S (respectively).
The material ejected between 738 and 1187 years ago (i. e., between
$t_1$ and $t_2$) has been ``processed'' by the working surface (i. e.,
it has gone through one of the two working surface shocks, and has either
piled up in the region between the two shocks or has been ejected
sideways).

Let us assume that the HH~34S working surface is the result of
a sinusoidal ejection velocity variability.
From the values of $v_1$ and $v_2$ it is then possible to estimate the
values of $v_0$ and $v_a$ (see equation 1) as~:

$$v_0\approx {{v_1+v_2}\over 2}=280\,{\rm km\,s}^{-1}\,,\eqno(13)$$

$$v_a\approx {{v_1-v_2}\over 2}=65\,{\rm km\,s}^{-1}\,.\eqno(14)$$

\noindent In order to obtain an estimate of the period of the
variability, we consider the fact that HH~34S has to be at
a distance from the source larger than the distance at which it
is formed. In other words,
we that $x_{34}> x_c$. From equations (2) and (3) we then have~:

$$\tau_a< {2\pi\,v_a\over {v_0}^2} x_{34}=1425\,{\rm yr}\,.\eqno(15)$$

\noindent Also, $\tau_a$ cannot be much smaller than this value,
since we would otherwise already see a steep velocity rise
along the jet as we go towards the source (corresponding to
a ``younger'' working surface being formed by this variability).
We therefore pick a value $\tau_a=1200$~yr.
In this way, we have obtained estimates for the three parameters
($v_0=280$~km~s$^{-1}$, $v_a=70$~km~s$^{-1}$ and $\tau_a=1200$~yr, see
equation 1) of a hypothetical sinusoidal ejection velocity variability
that gives rise to HH~34S.

\subsection{The other knots}

Let us first consider knots E-L. These knots are quasiperiodic,
in the sense that the separation $\Delta x$ between successive
knots is quite constant. Using the knot positions and velocities tabulated in
Table 1, we find a mean knot separation $<\Delta x>={1.71\times 10^{16}}$~cm,
and a mean spatial velocity $<v>=201$~km~s$^{-1}$. From equation (3)
we can then obtain the period $\tau_b$ of the variability required to produce
the observed knot structure~:

$$\tau_b\approx {{<\Delta x>}\over {<v>}}=27.0\,{\rm yr}\,.\eqno(16)$$

\noindent Now, if we assume that the knots correspond to working surfaces
in the asymptotic regime of large distances from the source, we can
compute the velocity jump across the successive working surfaces through
equation (5). Through this excercise we obtain velocity jumps in
the $\Delta v=15$-55~km~s$^{-1}$ range.

If we assume that the ejection velocity variability producing these
working surfaces is sinusoidal, it has to have a (half-)amplitude $v_b$
of the order of $\Delta v/2$. So, we have to pick a value of $v_b$
between 7 and 27~km~s$^{-1}$. For the numerical simulation described
in the following section we have picked a $v_b=15$~km~s$^{-1}$ value.

Finally, we consider knot O. This knot is placed at a distance
$x_O={4.2\times 10^{17}}$~cm from the source (see table 1), and does not
correspond to neither the variability mode of HH~34S nor the mode that
forms knots E-I. We also see evidence for a third mode in the modulation
of the flow velocity from knot E to L (see figure 3). A possible
estimate for the period $\tau_c$ of this third mode is~:

$$\tau_c\approx 2{{(x_O-x_G)}\over {(v_O+v_G)}}=310\,{\rm yr}\,.\eqno(17)$$

\noindent The amplitude $v_c$ of this mode can be estimated as~:

$$v_c\approx {{(v_G-v_K)}\over 2}\,,\eqno(18)$$

\noindent (see figure 3). If we use the velocity values from Table 1
(which correspond to de-projected proper motions, see above), we obtain
$v_c\approx 50$~km~s$^{-1}$. However, almost exactly one-half of this
value is obtained evaluating (18) with the de-projected radial velocities
of the knots. We therefore adopt a value $v_c= 40$~km~s$^{-1}$ for
the third mode of the ejection velocity variability.

\section{A numerical simulation}

We now describe the results of a numerical simulation of a jet
with a time-dependent ejection velocity given by~:

$$u_0(t)=v_0-v_a\,\sin \omega_a t\,+\,v_b\,\sin \omega_b t
\,+\,v_c\,\sin \omega_c t\,,\eqno(19)$$

\noindent with $v_0=280$~km~s$^{-1}$, $v_a=70$~km~s$^{-1}$,
$\tau_a=2\pi/\omega_a=1200$~yr (see section 3.3), $v_b=15$~km~s$^{-1}$,
$\tau_b=27$~yr, $v_c=40$~km~s$^{-1}$ and $\tau_c=310$~yr
(see section 3.4). We assume that the jet has a time-independent
injection density (of atoms and ions) $n_j=500$~cm$^{-3}$
and temperature $T_j=1000$~K, and that it travels into
a uniform environment of density $n_{env}=5$~cm$^{-3}$
and temperature $T_{env}=15$~K. Both the environment
and the injected beam are assumed to be neutral, with the
except of C and S which are assumed to be singly ionized.
The jet has an initial radius $r_j=10^{16}$~cm.

We should note that the initial radius that we have chosen for the
jet is larger than the radius of the HH 34 jet by a factor of $\sim 4$.
We have chosen such a larger radius in order to resolve the jet
diameter with $\sim 20$ points, which is already a quite limited
resolution. Also, the initial density that we have chosen for
the jet is too small probably by more than an order of magnitude.
This choice is again forced on us by the limited spatial resolution
of our numerical simulation (see below). With the lower density which
has been chosen, we can marginally resolve the recombination regions
behind the shocks in the flow, so that an approximate prediction
of the emitted spectrum can be carried out.

The cylindrically symmetric gasdynamic equations, together with rate
equations for up to six ions of H, He, C, N, O, Ne and S are integrated
in time with the adaptive grid Coral code. A detailed description of this
code, together with tests of the microphysical network, is
given by Raga, Mellema and Lundqvist (1997), Raga et al. (1997b),
and Mellema et al. (1998).

The cylindrical computational grid has an axial extent of
$10^{18}$~cm, a radial extent of $2.5\times 10^{17}$~cm,
and has a 4-level binary adaptive grid with a maximum
resolution of $9.75\times 10^{14}$~cm (in both the axial and
radial directions). The jet is injected from
the left side of the grid, and a reflection condition is applied on
the $x=0$ plane for radii greater than $r_j$. In our simulation,
the leading bow shock does not reach the outer radial boundary
of the grid.

An outflow boundary condition is applied at the right hand side
($x=10^{18}$~cm) of the computational grid. In our simulation,
the jet is allowed to propagate out of the computational
grid, in order to simulate the fact that the HH 34 jet is preceded
by previous ejection episodes (which form part of the ``superjet'',
see section 1).

In order to discuss the general characteristics of the flow,
in Figures 5 and 6 we show the pressure stratification and on-axis cuts
of the density, temperature and axial velocity corresponding to time
integrations of $t=1400$ and 2900 years, respectively. In the
$t=1400$~yr frame (figure 5), the leading working surface of the
jet is reaching the end of the computational grid. A series of
working surfaces of different strengths can be seen~:

\begin{itemize}

\item the leading working surface (at $x\approx 10^{18}$~cm),

\item two working surfaces
with velocity jumps of $\sim 100$~km~s$^{-1}$ (at $x\approx
6.5\times 10^{17}$~cm and $x\approx 7.5\times 10^{17}$~cm),

\item a number of ``small'' working surfaces (with velocity jumps
of $\sim 20$-50~km~s$^{-1}$) close to the injection point
(with $x<10^{17}$~cm).

\end{itemize}

\noindent As is clear from the axial velocity cut (bottom panel
of figure 5), the velocity jumps across the ``small'' working surfaces
rapidly decay and disappear for distances $x>10^{17}$~cm. However,
these working surfaces still can be seen at larger distances from
the source as clumps of enhanced density and temperature which cool
as they travel away from the source.

The two working surfaces
at $x\sim 10^{17}$~cm move at different velocities ($v_x\approx
320$~km~s$^{-1}$ for the $x\approx 6.5\times 10^{17}$~cm working
surface, and $v_x\approx 250$~km~s$^{-1}$ for the $x\approx 7.5\times
10^{17}$~cm working surface). Because of this difference in their
velocities, the two working surfaces merge into a single shock
structure before getting to the end of the computational grid.
Such working surface mergers are characteristic of multi-mode
variable ejection velocity models.

In Figure 6, we show the flow stratification corresponding
to a $t=2900$~yr time integration. The leading working
surface of the flow has of course left the computational grid
by now. However, the far wings of the leading bow shock can still
be seen as an outwardly propagating, almost cylindrical shock wave
(located at a radius $r\approx 1.5\times 10^{17}$~cm, see the top
panel of figure 6). Otherwise, the flow stratification is qualitatively
similar to the one found for the $t=1400$~yr time integration (see
figure 6).

We judge that the evolution close to the $t=2900$~yr time
frame is more appropriate for carrying out comparisons with the
HH 34 jet, since in the observations we have clear evidence (see
section 1) that HH~34S is not the leading working surface of the
outflow. In the following section we then discuss the time-sequence
of the flow around $t\sim 3000$~yr on the basis of the predicted
intensity maps.

\section{Comparisons between theory and observations}

In order to deduce the parameters for the ejection velocity variability
of our model, we have used the radial velocities, proper motions and
knot positions observed for the HH~34 jet. Therefore, by construction
our numerical model has knots with positions and kinematical
properties which agree well with the observations of HH~34.

However, for the determination of the model parameters we have not
used the observed line intensities. It is therefore of interest to compute
intensity maps from our numerical simulation, and to compare them with
images of HH~34. This comparison to some extent is an independent check
on the applicability of a variable ejection velocity model for this
object.

Figure 7 shows a time sequence of H$\alpha$ intensity maps computed
from the numerical simulation described in section 4. The maps have been
computed at intervals of $\Delta t=100$~yr, starting at an integration
time $t_0=2500$~yr. An angle of $30^\circ$ (as appropriate for HH 34, see
section 3) between the outflow axis and the plane of the sky has
been assumed, and the maps have not been convolved with a simulated
point spread function.
As is clear from Figure 7, our model produces images which are
qualitatively similar to several jet-like HH objects , e.~g.,
HH~111 (Reipurth 1989b; Reipurth et al. 1997) and HH~228 (the
outflow associated with Th 28, see Krautter 1986; Graham \& Heyer 1989).

In our model, the short period, low amplitude knots are modulated
into ``trains'' of knots by the intermediate period variability.
Also, the successive trains of knots have different intensities
as a result of the modulation by the long period variability mode.
As a result of this 3-mode interaction, quite different morphologies
are obtained as the time integration progresses, and the relative
phases of the three modes change.

From the time sequence of Figure 7, we choose the $t=2900$~yr frame.
In Figure 8, we show a comparison of the predicted H$\alpha$ and
[S~II]~6717+6731 maps for this time frame with the corresponding
observations of the HH~34 jet. From this comparison, it is clear
that our numerical model produces intensity maps which are qualitatively
quite similar to the ones of HH~34.

In order to quantify this result, in Figure 9 we have plotted the observed
and predicted H$\alpha$ luminosities of the consecutive knots as a function
of de-projected distance from the source. The most important difference
between the predictions and the observations is that the predicted
luminosities are lower than the observed ones by a factor of $\sim 100$.
This result indicates that the initial number density that we chose for
the jet in our simulation ($n_j=500$~cm$^{-3}$, see section 4) is
probably too low by a factor of $\sim 100$ with respect to the
initial density of the HH~34 jet.

If we multiply the predicted luminosities by a factor of 100
(see figure 9), we
find that the predictions match the observed luminosities of the
successive knots along the HH~34 jet to within a factor of $\sim 10$.
Given the extremely strong dependence of the predicted luminosities
on shock velocity (see, e.~g., Hartigan, Raymond and Hartmann 1987;
Raga \& Kofman 1992), this agreement is nothing short of surprising.

With the initial jet density ($n_j\approx {5\times 10^4}$~cm$^{-3}$)
necessary to reproduce the observed luminosities, and the initial jet radius
$r_j=10^{16}$~cm and average velocity $v_0=280$~km~s$^{-1}$ of our
simulated jet (see section 4), we can compute an average mass loss
rate of ${\dot M}\approx {1.6\times 10^{-5}}$~M$_\odot$yr$^{-1}$ for
the HH 34 jet.

It would of course be interesting to carry out a comparison between,
e.~g., the predicted morphologies and line ratios for the successive
knots and the corresponding observations. However, our numerical simulation
does not have a high enough spatial resolution for such comparisons to
be meaningful.

\section{Conclusions}

We have first reviewed the analytic models for the formation and propagation
of internal working surfaces, and derived a simple analytic description
of the interaction between different oscilation modes of the ejection
velocity variability. We find that multi-mode ejection velocity
variabilities lead to the formation of ``trains'' of knots travelling
away from the source.

Using these analytic models, we have proceeded
to analyze the observations of the HH~34 jet which have been
obtained in the past. Through this analysis, we have derived three
modes for an assumed ejection velocity variability of the source
of HH~34.

These modes have periods and (half-)amplitudes~:

\begin{itemize}

\item $\tau_a=1200$~yr, $v_a=70$~km~s$^{-1}$,

\item $\tau_b=27$~yr, $v_b=15$~km~s$^{-1}$,

\item $\tau_c=310$~yr, $v_c=40$~km~s$^{-1}$,

\end{itemize}

\noindent and a $v_0=280$~km~s$^{-1}$ average velocity.
If we plot these periods and amplitudes logarithmically, it is clear
that they closely follow a $v\propto \tau^{0.4}$ power law (see
figure 10). Due to the uncertainties in the determination of these
parameters, however, this result should probably be taken as a mere
curiosity.

We have then proceeded to carry out a numerical simulation of a jet
with this ejection velocity variability. From this simulation, we
obtain predicted H$\alpha$ (and [S~II] 6717+6731) intensity maps which
are qualitatively very similar to the corresponding observations of
the HH~34 jet. In order to reproduce the luminosities of the observed
knots, we find that we
would need an initial number density for the jet which is
2 orders of magnitude larger than the one we have used. However, with
the resolution of our simulation, we cannot use such a high density
(as the cooling regions would then be completely unresolved).

From our model fit to the HH~34 jet, we obtain an estimate of
${\dot M}\approx {1.6\times 10^{-5}}$~M$_\odot$yr$^{-1}$ for this
outflow. Interestingly, this estimate is about an order of magnitude
larger than the mass loss rate computed by Raga (1991) on the basis
of a comparison with a steady jet model, and about three orders
of magnitude higher than the less elaborate estimate of Mundt,
Brugel and B\"uhrke (1987). These differences are a good illustration
of the fact that mass loss rate estimates for HH jets are very
highly model dependent.

\acknowledgments
The work of A. Raga was supported by grants from DGAPA, CONACyT and
the UNAM/Cray program. We would like to thank Bo Reipurth and Steve Heathcote
for providing us with their ESO NTT images of HH~34.
Also, we would like to thank Dave Devine, Bo Reipurth and John Bally
for providing us with their images of the HH 34 superjet.

\begin{deluxetable}{cccc}
\scriptsize
\tablewidth{0pc}
\tablenum{1}
\tablecaption{Parameters of the knots along the HH 34 jet}
\tablehead{Knot & $x$\tablenotemark{1}~(cm) & $L(L_\odot)$ &
$v$\tablenotemark{2}~(km~s$^{-1}$)}
\startdata
E &   ${9.64\times 10^{16}}$ & ${1.46\times 10^{-5}}$ & 201 \\
F &   ${1.12\times 10^{17}}$ & ${1.10\times 10^{-5}}$ & 256 \\
G &   ${1.31\times 10^{17}}$ & ${1.09\times 10^{-5}}$ & 288 \\
I &   ${1.50\times 10^{17}}$ & ${2.08\times 10^{-5}}$ & 209 \\
J &   ${1.63\times 10^{17}}$ & ${1.32\times 10^{-5}}$ & 209 \\
K &   ${1.84\times 10^{17}}$ & ${7.82\times 10^{-6}}$ & 184 \\
L &   ${2.16\times 10^{17}}$ & ${2.34\times 10^{-6}}$ & 240 \\
O &   ${4.20\times 10^{17}}$ & ${1.42\times 10^{-5}}$ & 318 \\
34S & ${8.01\times 10^{17}}$ & ${2.26\times 10^{-3}}$ & 329 \\
87  & ${5.14\times 10^{18}}$ & ${9.24\times 10^{-4}}$ & \phantom{0}80 \\
88  & ${5.38\times 10^{18}}$ & ${3.08\times 10^{-4}}$ & \phantom{0}50 \\
\tablenotetext{1}{de-projected distances from the source to knot center}
\tablenotetext{2}{de-projected proper motion velocities, except
for knots 87,88 for which de-projected radial velocities are given}
\enddata
\end{deluxetable}
\clearpage

\vfill\eject

\figcaption{Diagram showing the time-dependent ejection velocity
$u_0(t)$ for two-mode variabilities. The graphs show the results for
two different combinations of parameters (listed below each panel),
which are discussed in the text.
\label{fig1}}

\figcaption{H$\alpha$ and [S~II]~6717+6731 HST images of the region close
to the source of HH~34. The images are shown with a logarithmic greyscale
(covering different ranges for each of the images, as shown by the
wedges to the right of the panels). The region within $10''$ of the source
shows a number of very faint and narrow knots, which are seen more distinctly
in the [S~II] image.
\label{fig2}}

\figcaption{Spatial velocities (top) and H$\alpha$ luminosities (bottom)
as a function of the de-projected distance $x$ from the source of HH~34.
The dashed lines in the bottom panel show the $L_{H\alpha}\propto x^{-4.8}$
evolution predicted for the knots from the asymptotic solution for large
distances from the source. The H$\alpha$ luminosities have been obtained
from measurements on images kindly provided to us by Heathcote, Reipurth,
Devine and Bally.
\label{fig3}}

\figcaption{Schematic representation of the position-velocity diagram of
HH~34S of Heathcote \& Reipurth (1992). The slit was aligned with the
outflow axis, and the direction away from the outflow source is towards
the right. The velocity of the surrounding cloud is shown with a solid,
horizontal line. The working surface is seen to be overruning weakly
emitting material with a (de-projected) velocity $v_2=214$~km~s$^{-1}$.
The jet is feeding into the working surface from the upstream direction
with a (de-projected) velocity $v_1=344$km~s$^{-1}$.
\label{fig4}}

\figcaption{Flow stratification obtained from the numerical simulation
described in the text for a $t=1400$~yr time integration. The top
panel shows the pressure stratification of the flow (with factor of
2 contours). The three bottom panels show number density (of atoms
and ions), temperature and axial velocity cuts down the symmetry
axis of the flow.
\label{fig5}}

\figcaption{Flow stratification obtained from the numerical simulation
described in the text for a $t=2900$~yr time integration. The top
panel shows the pressure stratification of the flow (with factor of
2 contours). The three bottom panels show number density (of atoms
and ions), temperature and axial velocity cuts down the symmetry
axis of the flow.
\label{fig6}}

\figcaption{Logarithmic greyscale representation of the H$\alpha$
intensity maps predicted from the numerical simulation described in
the text for time integrations $t=2500$ (top panel), 2600, 2700, 2800, 2900
and 3000~yr (bottom panel). The maps have been normalized so that the
peak intensity in the leading working surface of the $t=2900$~yr panel
has a value of 1.
\label{fig7}}

\figcaption{Comparison of the H$\alpha$ and [S~II]~6717+6731 images predicted
from the numerical simulation described in the text for a $t=2900$~yr
time integration and images of the HH~34 outflow in the corresponding
lines. The images have been scaled arbitrarily, and a constant background
has been subtracted from the observations. The colours represent a
logarithmic scale, covering the same range for the predicted and the
correponding observed images. A smaller range has been used for the
H$\alpha$ images, since the observed H$\alpha$ image has a smaller
dynamic range (see the wedges on the right-hand side of the plots).
The predicted images have the same spatial coverage as the
frames of Figure 7. The HH~34 images have been scaled in
an approximate way, assuming a distance of 450~pc.
The observations of HH~34 have been previously published by
Heathcote \& Reipurth (1992), who we thank for kindly having provided
us with the images.
\label{fig8}}

\figcaption{H$\alpha$ luminosities of the knots along HH~34 (squares)
as a function of de-projected distance from the source (also see table 1).
The luminosities predicted for the knots seen in the $t=2900$~yr H$\alpha$
image from the numerical simulations are plotted (after having been
multiplied by a factor of 100) as crosses.
\label{fig9}}

\figcaption{Logarithmic plot of the velocity half-amplitude of the
three modes (used to model HH~34) as a function of their period. The
straight line corresponds to a $v\propto v^{0.4}$ power law.
\label{fig10}}

\clearpage
 
\begin{figure}[v]
\plotone{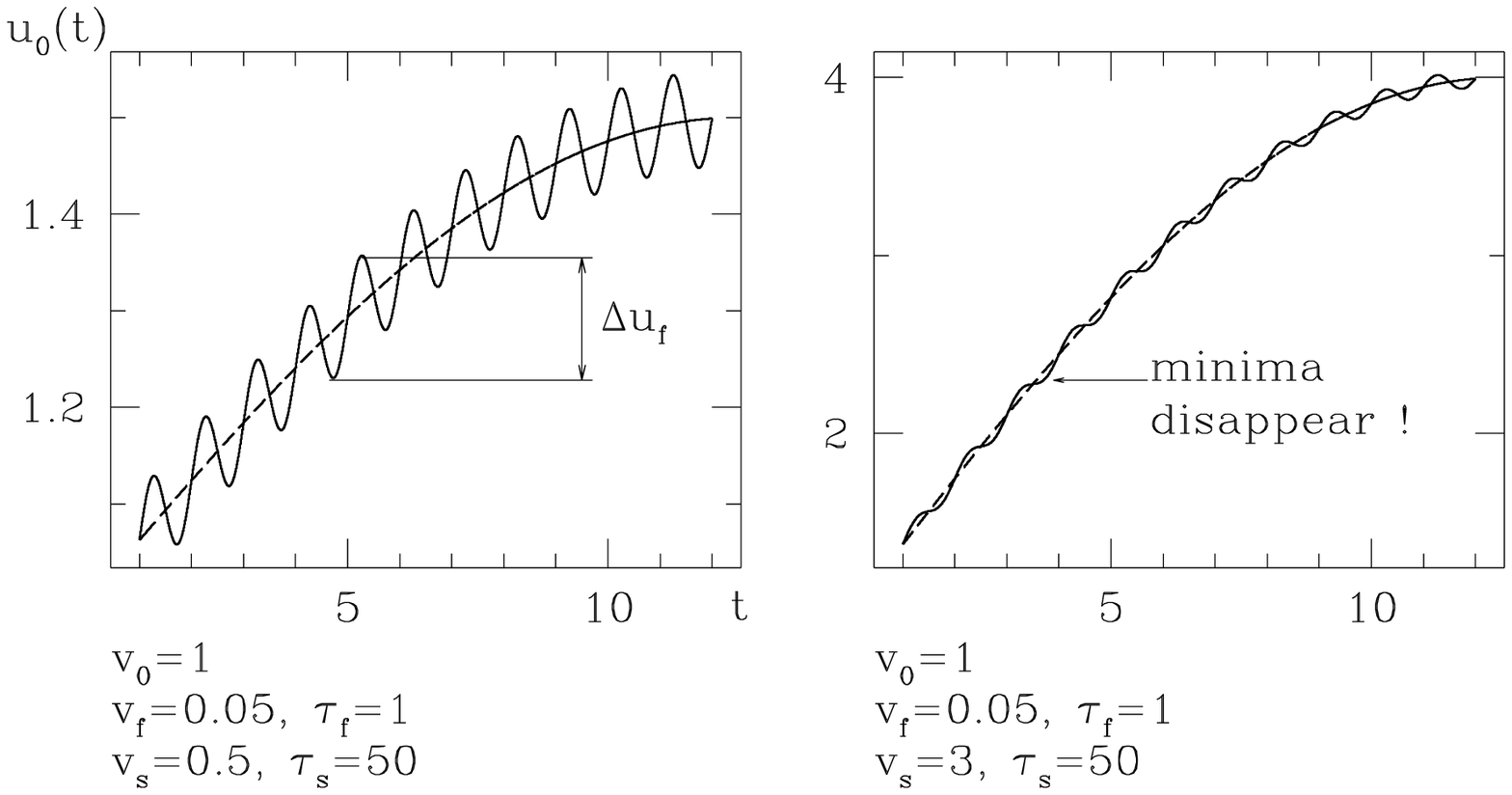}
\end{figure}

\clearpage
\begin{figure}[v]
\plotone{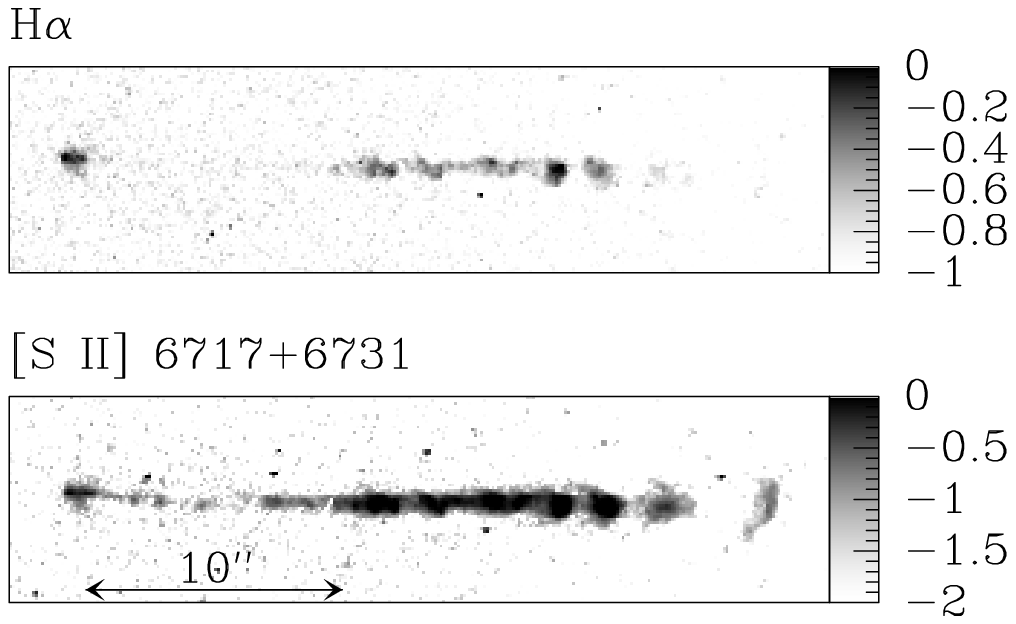}
\end{figure}

\clearpage
\begin{figure}[v]
\plotone{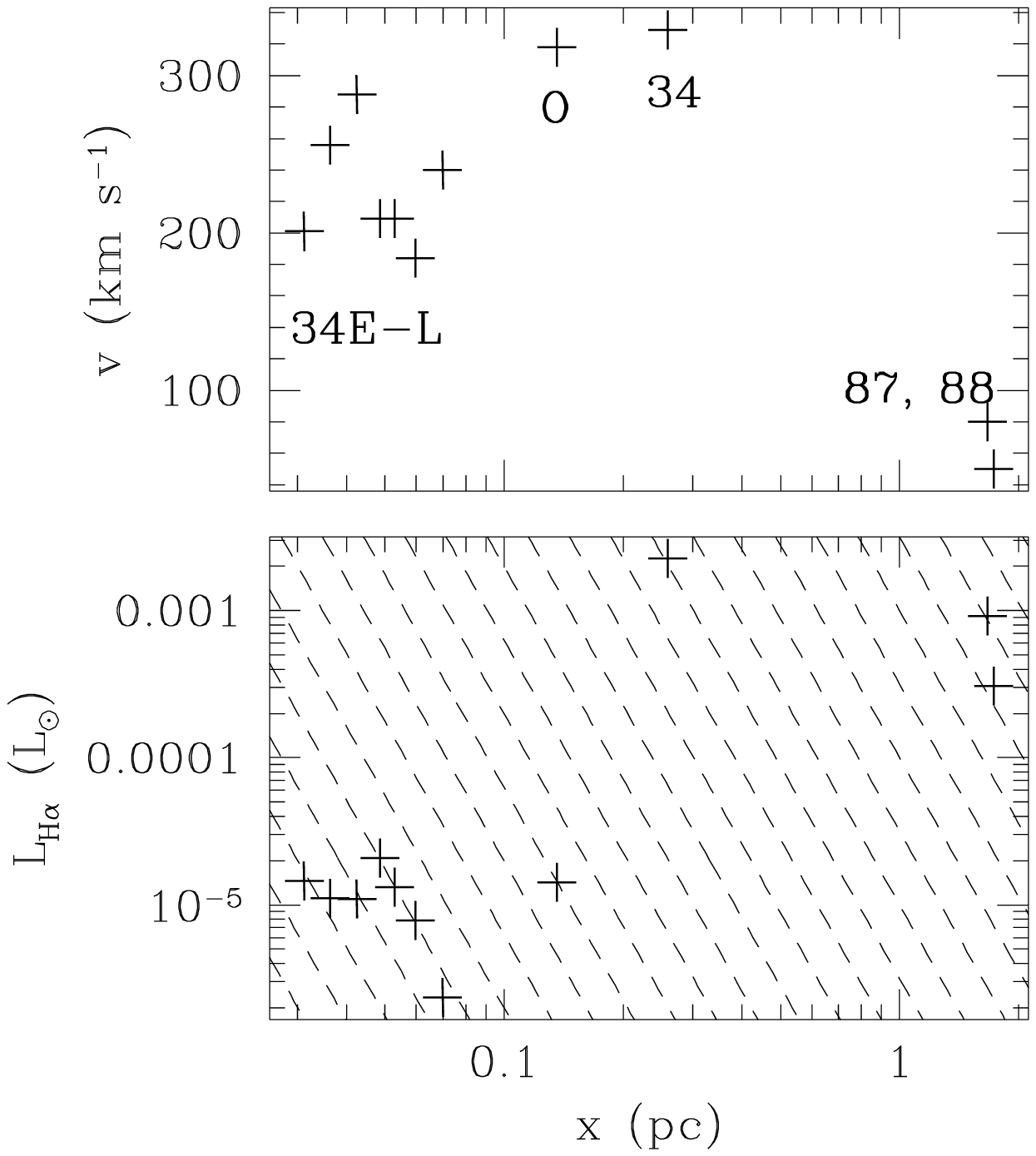}
\end{figure}

\clearpage
\begin{figure}[v]
\plotone{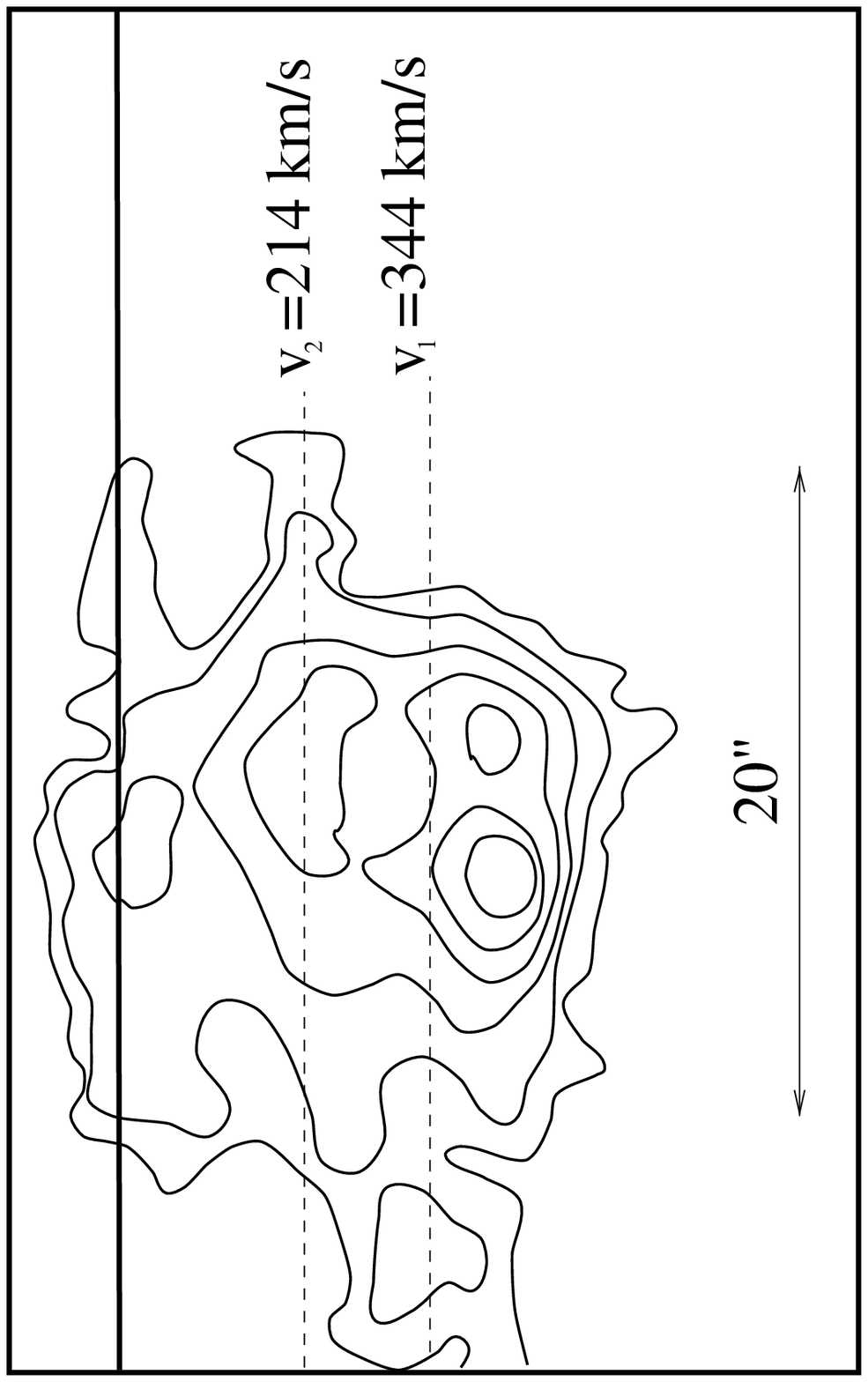}
\end{figure}

\clearpage
\begin{figure}[v]
\plotone{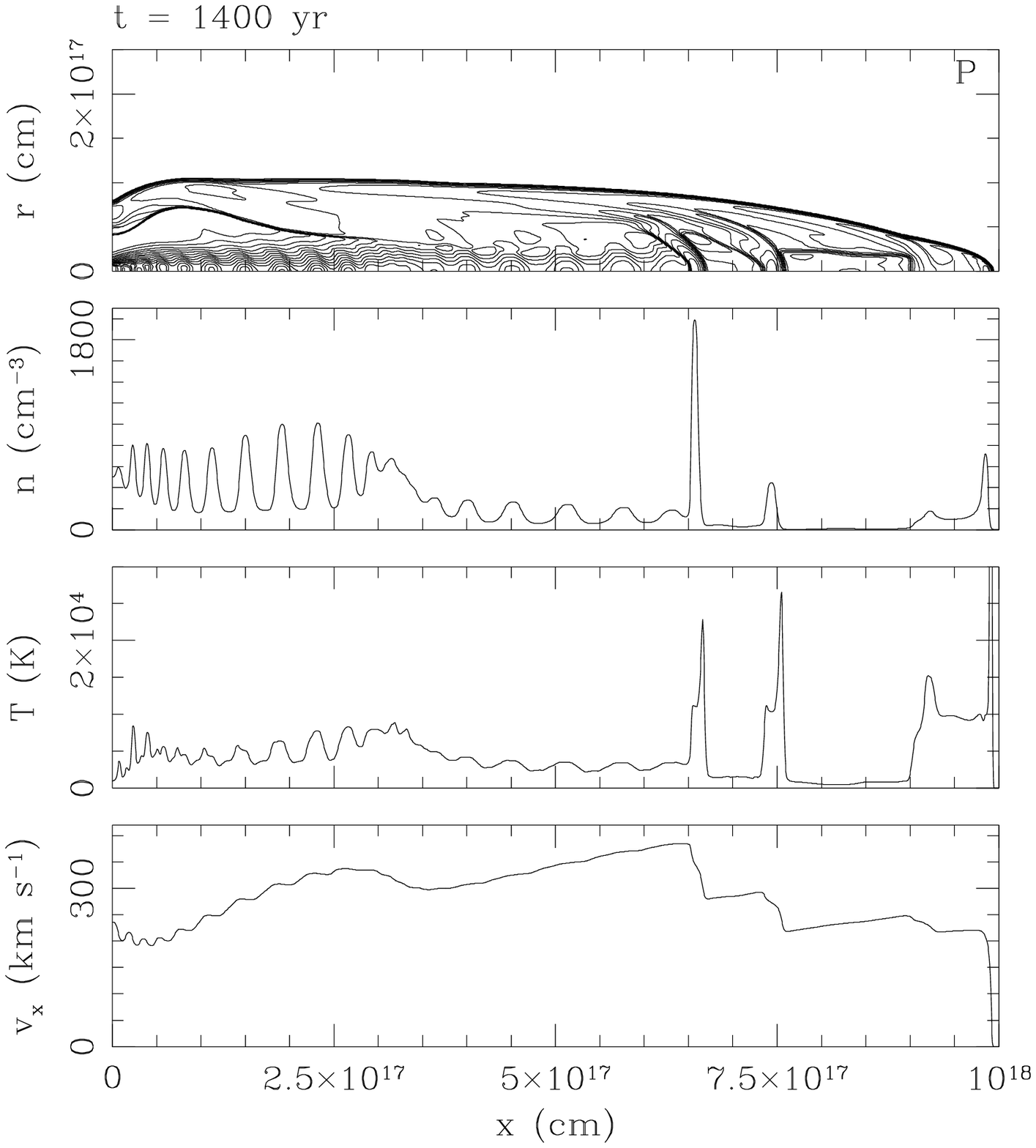}
\end{figure}

\clearpage
\begin{figure}[v]
\plotone{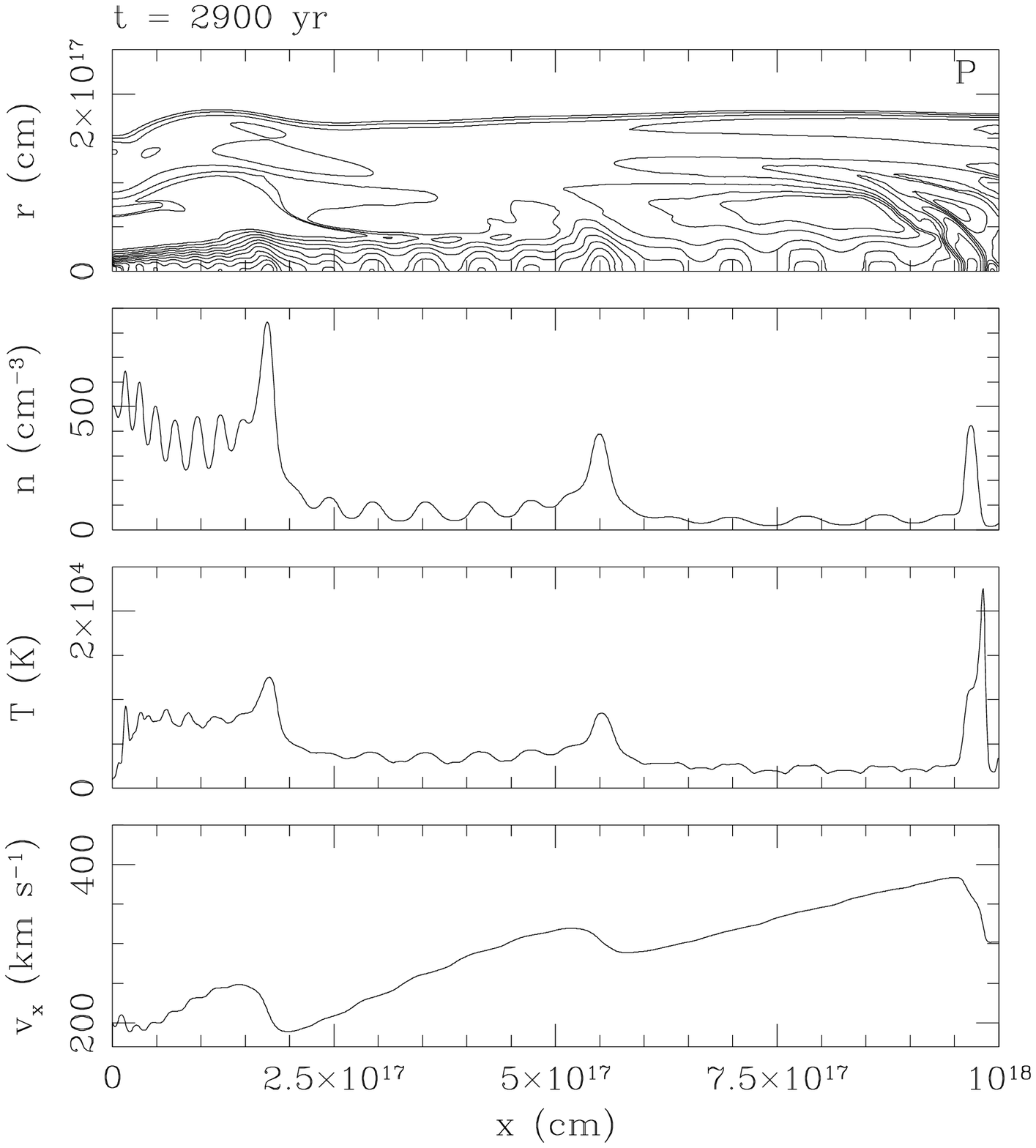}
\end{figure}

\clearpage
\begin{figure}[v]
\plotone{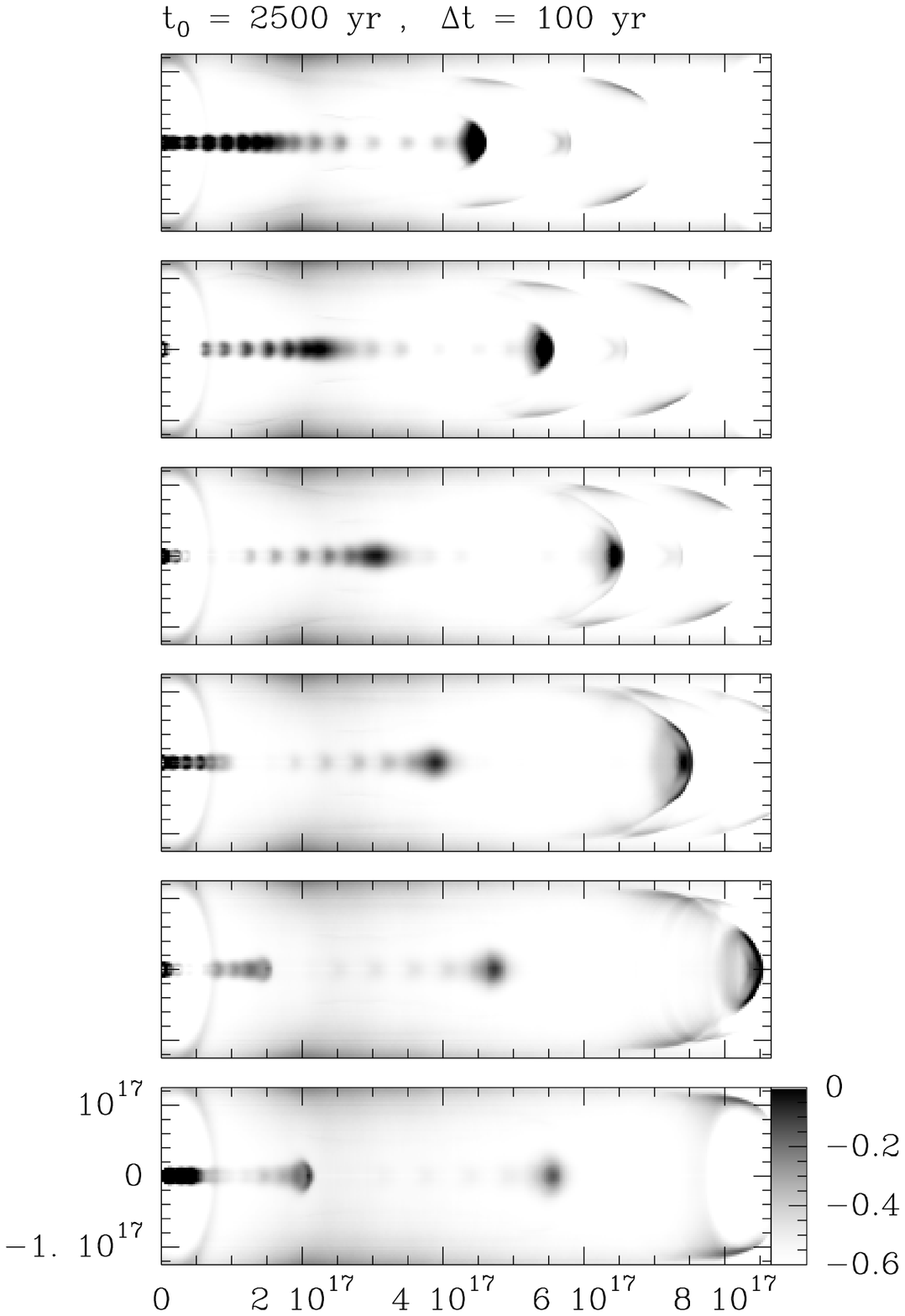}
\end{figure}

\clearpage
\begin{figure}[v]
\plotone{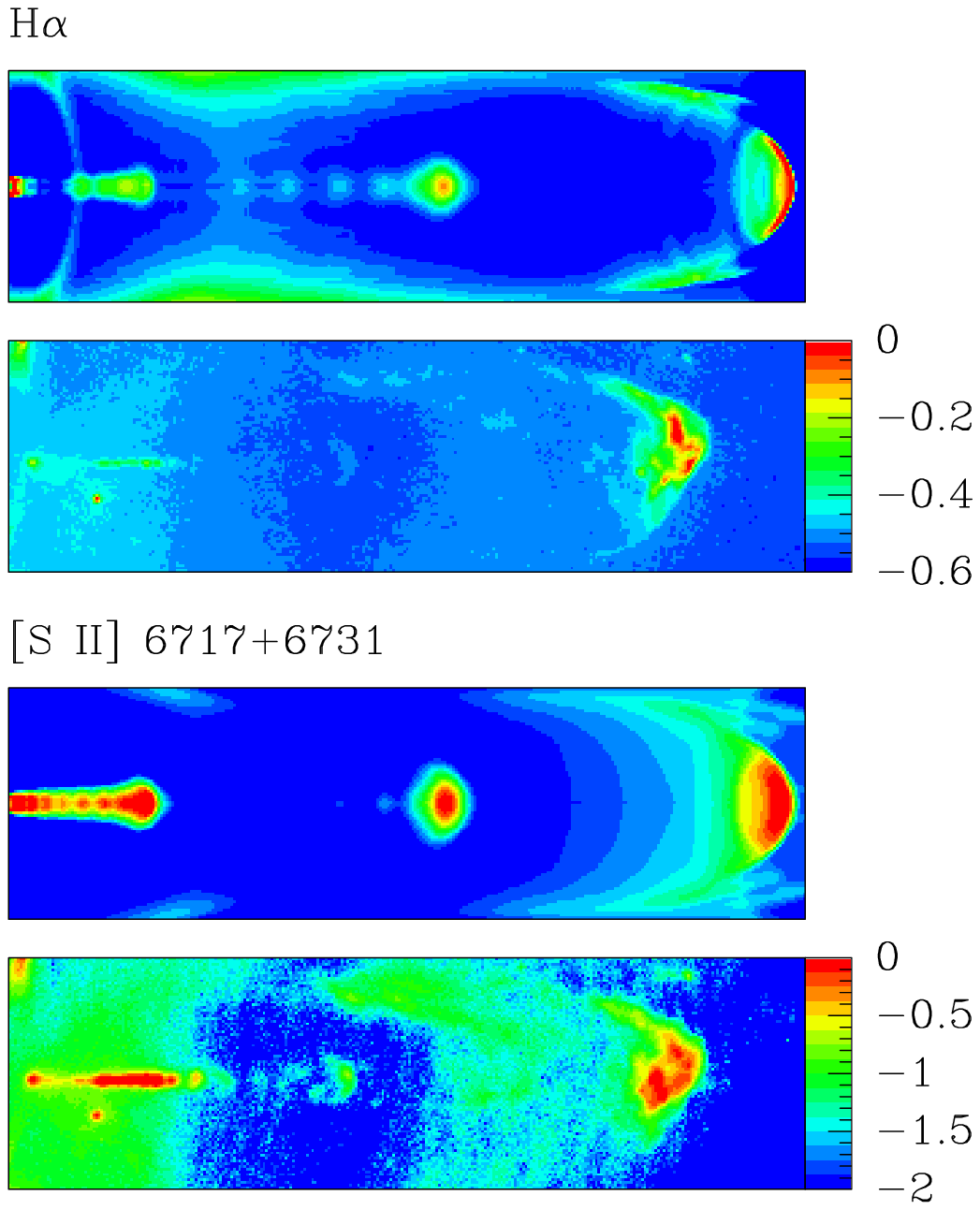}
\end{figure}

\clearpage
\begin{figure}[v]
\plotone{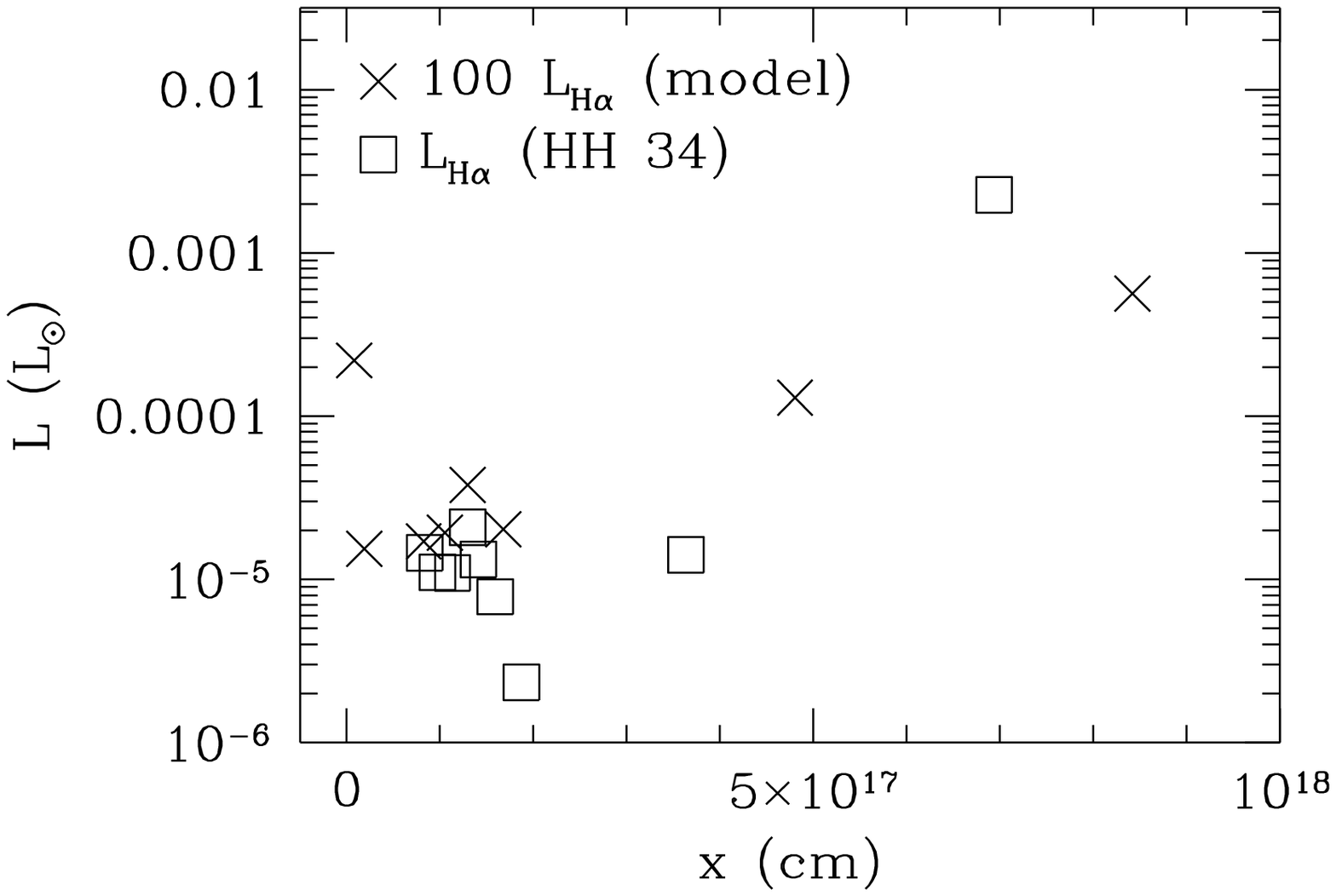}
\end{figure}

\clearpage
\begin{figure}[v]
\plotone{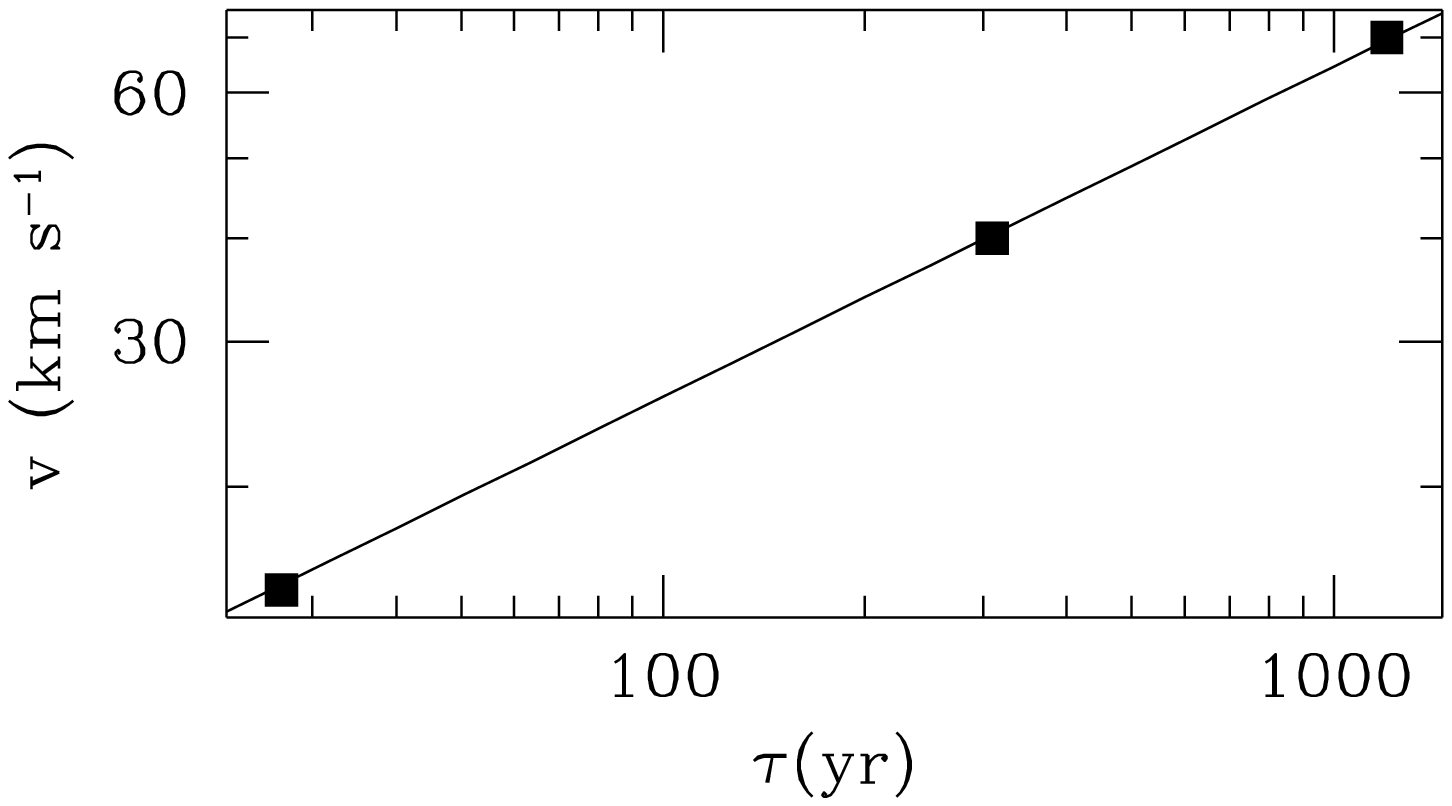}
\end{figure}

\clearpage

\end{document}